\documentclass[letter,twocolumn,aps,nofootinbib]{revtex4-1}

\usepackage{epsfig}
\usepackage{amsmath}
\usepackage{amssymb}
\usepackage{verbatim}
\usepackage{bm}
\usepackage{dsfont}
\usepackage{subfigure}
\usepackage{cancel}
\usepackage{color}
\usepackage[utf8]{inputenc} 

 \makeatletter
    \renewcommand\@make@capt@title[2]{%
     \@ifx@empty\float@link{\@firstofone}{\expandafter\href\expandafter{\float@link}}%
      {\textbf{#1}}\@caption@fignum@sep#2\quad}%
    \makeatother
   
\makeatletter 
\renewcommand{\fnum@figure}{\textbf{Figure~\thefigure}}
\makeatother

\begin{document}

\begin{flushleft}
ULB-TH/16-01 \\ IFT-UAM/CSIC-16-004 \\ CPHT-RR001.012016 \\ \textcolor{white}{blank space} \\ \textcolor{white}{blank space}
\end{flushleft}

\title{UV Corrections in Sgoldstino-less Inflation}

\author{Emilian Dudas}
\affiliation{Centre de Physique Th\'eorique, Ecole Polytechnique, CNRS, Univ. Paris-Saclay, 91128 Palaiseau Cedex, France}
\author{Lucien Heurtier}
\affiliation{Service de Physique Th\'eorique, Universit\'e Libre de Bruxelles, Boulevard du Triomphe, CP225, 1050 Brussels, Belgium}
\author{Clemens Wieck}
\affiliation{Departamento de F\'isica Te\'orica UAM and Instituto de F\'isica Te\'orica UAM/CSIC, Universidad Aut\'onoma de Madrid Cantoblanco, 28049 Madrid, Spain}
\author{Martin Wolfgang Winkler}
\affiliation{Bethe Center for Theoretical Physics and Physikalisches Institut der Universit\"at Bonn, \\
Nussallee 12, 53115 Bonn, Germany}

\begin{abstract}
We study the embedding of inflation with nilpotent multiplets in supergravity, in particular the decoupling of the sgoldstino scalar field. Instead of being imposed by hand, the nilpotency constraint on the goldstino multiplet arises in the low energy-effective theory by integrating out heavy degrees of freedom. We present explicit supergravity models in which a large but finite sgoldstino mass arises from Yukawa or gauge interactions. In both cases the inflaton potential receives two types of corrections. One is from the backreaction of the sgoldstino, the other from the heavy fields generating its mass. We show that these scale oppositely with the Volkov-Akulov cut-off scale, which makes a consistent decoupling of the sgoldstino nontrivial. Still, we identify a parameter window in which sgoldstino-less inflation can take place, up to corrections which flatten the inflaton potential.
\end{abstract}

\maketitle

%

\section{Introduction}

Constrained chiral multiplets or, equivalently, nilpotent superfields and their application to cosmology have attracted a large amount of interest in recent years \cite{AlvarezGaume:2010rt,Achucarro:2012hg,Antoniadis:2014oya,Buchmuller:2014pla,Ferrara:2014kva,Kallosh:2014via,Dall'Agata:2014oka,Kallosh:2014hxa,Linde:2015uga,Carrasco:2015uma,Kahn:2015mla,Scalisi:2015qga,Carrasco:2015pla,Dudas:2015eha,Aparicio:2015psl,Ferrara:2015tyn,Carrasco:2015iij}. One feature of theories with nonlinear supersymmetry, i.e., with a constrained multiplet satisfying $S^2 = 0$, is the absence of a dynamical scalar degree of freedom. The auxiliary field of $S$ breaks supersymmetry and the goldstino fermion is the only propagating field \cite{Volkov:1973ix,Rocek:1978nb,Ivanov:1978mx,Lindstrom:1979kq}. This makes them appealing in cosmological model building for various reasons. 

The connection of such theories to string theory has recently been studied in \cite{McGuirk:2012sb,Kallosh:2014wsa,Bergshoeff:2015jxa,Kallosh:2015nia,Bandos:2015xnf,Aparicio:2015psl,Garcia-Etxebarria:2015lif}. Effective supergravity theories with a constrained goldstino multiplet can be shown to arise from $\overline{D3}$-branes in certain geometries \cite{Kallosh:2014wsa,Bergshoeff:2015jxa,Kallosh:2015nia,Bandos:2015xnf,Garcia-Etxebarria:2015lif}. The emergence of nonlinear supersymmetry in string models with anti-branes was proven in \cite{Dudas:2000nv}, in the context of
global string theory vacua \cite{bsb}. In such UV embeddings it is difficult to extract the behavior of the supergravity above the cut-off scale of the Volkov-Akulov action. Usually there is no scale at which linear supersymmetry is restored and therefore the scalar component of $S$ does not exist. A step towards understanding the connection between the linear and nonlinear regimes was recently made in \cite{Kallosh:2015pho}, where it was shown explicitly that nonlinear supergravity theories are equivalent to linear supergravities with an infinitely heavy sgoldstino scalar, and that the limit relating the two is well-defined via functional integration. With a few restrictions this connection was previously known in the rigid limit \cite{Komargodski:2009rz}.\footnote{For a recent study regarding the applicability of nilpotency conditions cf.~\cite{Ghilencea:2015aph}.}

Therefore it is desirable to study field theory examples in which a heavy sgoldstino exists so that supersymmetry becomes linearly realized at a high scale. In such cases, the sgoldstino field cannot be infinitely heavy. Its mass, and hence the Volkov-Akulov cut-off scale, must be lower than the Planck scale -- and favorably below the Kaluza-Klein and string scales. A stronger constraint arises from unitarity which signals a perturbative breakdown of the nonlinear theory at a scale $\sim\sqrt{m_{3/2}}$ in Planck units. A UV complete theory which can describe both the linear and nonlinear regimes is bound to yield corrections which are missed by simply imposing a nilpotency constraint on the goldstino multiplet in supergravity. In this letter we compute these corrections and evaluate their effects in simple inflation models previously studied in the literature. It is our aim to prove that in a well-defined regime of the theory, corrections are under control -- though in a quite constrained parameter space.

For this purpose the class of models developed in \cite{Dall'Agata:2014oka} is particularly instructive.\footnote{We recommend \cite{Ferrara:2015cwa} as a review of these and other inflation models involving nonlinear supersymmetry.} They feature the coupling of a nilpotent stabilizer multiplet to a holomorphic function of the inflaton multiplet, giving rise to a plethora of possible potential shapes for the inflaton scalar. During inflation and in the vacuum supersymmetry is broken by the auxiliary field of the nilpotent multiplet. The setup can accommodate low-energy supersymmetry which is nontrivial given the high scale of inflation. We extend this setup to a  supergravity with heavy fields in which a large mass for the sgoldstino scalar is generated dynamically. We expect our results to be relevant in many other supergravity theories with nilpotent goldstino multiplets. Thus, we hope that this work is another step towards understanding nilpotent multiplets and their role in cosmology.

%

\section{Sgoldstino decoupling}\label{sec:sgoldstinodecoupling}

The success of nilpotent fields in cosmology has triggered growing interest in their field-theoretical origin. It is well-known that in spontaneously broken linear supersymmetry, the sgoldstino field acquires a large mass through the operator
\begin{align}\label{eq:quartk}
K\supset c\,\frac{|S|^4}{\Lambda^2}
\end{align}
in the K\"ahler potential. In the limit $c\rightarrow \infty$, the sgoldstino becomes infinitely heavy and the resulting theory is equivalent to nonlinearly realized supersymmetry with a nilpotent goldstino multiplet $S$ \cite{Kallosh:2015pho,Komargodski:2009rz}. Clearly this theory is only a low-energy effective theory.  With the sgoldstino decoupled, it violates perturbative unitarity at the intermediate energy scale $\sqrt{m_{3/2}}$~\cite{Casalbuoni:1988sx}. Requiring inflation in the perturbative regime one obtains the generic constraint \cite{Dall'Agata:2014oka}
\begin{align}\label{eq:unitarity}
m_{3/2} > H^2\,,
\end{align}
where $H$ denotes the Hubble scale. However, the scale of supersymmetry breaking may be different during and after inflation. Hence, nilpotent inflation models consistent with low-energy supersymmetry can be constructed~\cite{Dall'Agata:2014oka}.

A different concern is the limit $c\rightarrow \infty$: in a UV-complete model the operator \eqref{eq:quartk} arises from couplings of $S$ to heavy degrees of freedom.  As an example, we may consider the superpotential coupling $W \supset \lambda S X^2$ of $S$ to the heavy field $X$ with mass $m_X$. This coupling generates a one-loop correction \cite{Grisaru:1996ve},
\begin{align}
 K \supset -\frac{\lambda^4}{16\pi^2}\frac{|S|^4}{m_X^2}\;.
\end{align}
The limit $c\rightarrow\infty$ in~\eqref{eq:quartk} then corresponds to taking the coupling $\lambda$ to infinity or the mass $m_X$ to zero. Since both must be finite and $m_X$ must be large for the effective field theory (EFT) to make sense, we must consider the regime where the sgoldstino has a finite mass, i.e., finite $c$. In the remainder of this letter we strive to determine whether inflation is still possible in this case. Specifically, we determine whether the inflaton potential obtained in the nilpotent limit still holds and corrections are under control. 

We will find that such corrections are of two different natures. Additional heavy fields at the energy scale $\Lambda$ backreact on the inflaton potential, introducing corrections which vanish as $\Lambda \to \infty$.\footnote{These we call ``UV corrections'' because they arise from embedding the nilpotent multiplet in a complete theory of supergravity.} On the other hand, the finite mass of the sgoldstino field leads to corrections which vanish in the limit where the latter is infinitely heavy. This corresponds to $\Lambda \to 0$. Therefore, it is far from obvious that both types of corrections can be suppressed simultaneously.

Note that while we study this in the class of inflation models proposed in \cite{Dall'Agata:2014oka}, our findings can straightforwardly be applied to alternative scenarios with nilpotent multiplets.

%

\section{Sgoldstino-less models of inflation}\label{sec:sgoldstinoless}

Let us briefly review the inflation models of~\cite{Dall'Agata:2014oka}. They feature the K\"ahler and superpotential
\begin{subequations}
\begin{align}
K &=\frac{1}{2} (\Phi+\overline{\Phi})^2 +|S|^2\,, \\
W &= f(\Phi) \left(1+\sqrt{3} S\right)\,,
\end{align}
\end{subequations}
where $\Phi$ denotes the inflaton superfield and $S$ contains the stabilizer field. This setup is a generalization of the models developed in \cite{Kawasaki:2000yn,Kallosh:2010xz} with built-in supersymmetry breaking by the auxiliary field of $S$. The function $f$ satisfies $f(0)\neq 0$, $f'(0)=0$ and $\overline{f(x)}=f(-\bar{x})$. In \cite{Dall'Agata:2014oka} it is assumed that $S$ fulfills the boundary condition $S^2=0$ of a nilpotent chiral multiplet. This implies $\langle s \rangle =0$ for its scalar component.\footnote{We use capital letters for superfields and small letters for their scalar components.}

The factor $\sqrt3$ ensures the cancellation of the cosmological constant in the vacuum at $\langle \phi\rangle =0$. Along the inflationary trajectory the potential reads
\begin{align}
V = \left| f'\left(i\frac{\varphi}{\sqrt{2}}\right)\right|^2\,,
\end{align}
where $\varphi=\sqrt{2}\,\text{Im}\,\phi$ denotes the canonically normalized inflaton. Two examples for $f$ are discussed in \cite{Dall'Agata:2014oka}. One is 
\begin{align}
f(\Phi)=f_0 - \frac{m}{2}\Phi^2\,, \label{eq:chaotic}
\end{align}
leading to the potential of chaotic inflation, $V=\frac{1}{2}m^2\varphi^2$. The other is
\begin{align}
f(\Phi)= f_0 - i \sqrt{V_0} \left(\Phi+i\frac{\sqrt{3}}{2} e^{2i\Phi/\sqrt{3}}\right),
\end{align}
producing the plateau potential $V= V_0 \left( 1- e^{-\sqrt{2/3}\,\varphi}\right)$.\footnote{As pointed out in Section 5 of \cite{Dall'Agata:2014oka}, the function $f$ can be extended to include matter fields like an MSSM sector. Tachyonic directions are avoided automatically for matter fields which appear at least quadratically in $f$.}

In the following we call $S$ the goldstino multiplet and $s$ the sgoldstino, its scalar component. This is because $s$ is the heavy scalar that is supposed to decouple, and despite the fact that the inflaton multiplet has a sub-dominant but nonvanishing auxiliary field during inflation.

%

\section{Corrections from the sgoldstino}~\label{sec:sgold}

Let us discuss corrections to the inflaton potential which arise if the sgoldstino has a finite mass. To this end we consider
\begin{subequations}\label{eq:sugramodel}
\begin{align}
W &= f(\Phi) (1 + \delta  S )\,,\\
K &=\frac{1}{2} (\Phi+\overline{\Phi})^2 +|S|^2 - \frac{|S|^4}{\Lambda^2}\,.
\end{align}
\end{subequations}
The difference compared to the previous section is that we do not impose the nilpotency constraint $S^2=0$. Instead we introduce the term $|S|^4/ \Lambda^2$ in the K\"ahler potential which generates a large -- but finite -- mass for the sgoldstino $s$ and dynamically keeps $s$ close to the origin.\footnote{Compared to Section~\ref{sec:sgoldstinodecoupling} we absorbed the parameter $c$ in the definition of $\Lambda$.} Supersymmetry breaking introduces an inflaton-dependent linear term for the stabilizer field which slightly shifts it away from the origin \cite{Buchmuller:2014pla}. As this effect scales inversely with the mass of $s$, it is absent in the nilpotent limit. Notice that we introduced the parameter $\delta$ which allows us to tune the vacuum energy to zero at the minimum of the potential. Due to the shift of $s$, $\delta$ is close to but not exactly $\sqrt{3}$. We find
\begin{align}
\delta = \sqrt{3} + \frac{\Lambda^2}{2\sqrt{3}} + \mathcal{O}(\Lambda^4)\,.
\end{align}
For a compact notation we introduce
\begin{align}
V_0 = \left| f'\left(i\frac{\varphi}{\sqrt{2}}\right)\right|^2\,,
\end{align}
denoting the inflaton potential in the limit where $S$ is nilpotent and hence $s$ is infinitely heavy. The gravitino mass along the inflationary trajectory can be approximated as
\begin{align}
m_{3/2}^2 = e^K |W|^2 \simeq \left| f\left(i\frac{\varphi}{\sqrt{2}}\right)\right|^2\,.
\end{align}

As only the real part of the stabilizer field is displaced during inflation, we set $\bar{s}=s$ in the following. At second order in $s$ the scalar potential reads
\begin{align}\label{eq:sgoldstinopotential}
V = V_0 + m_{3/2}^2 \Lambda^2 + \sqrt{3} \left(2  V_0 - 4 m_{3/2}^2\right) s + m_s^2 s^2 \,,
\end{align}
including only terms up to $\mathcal{O}(\Lambda^2)$.\footnote{Notice that $s = \mathcal{O}(\Lambda^2)$ and $m_s^2=\mathcal{O}(\Lambda^{-2})$.}
The sgoldstino mass is given by 
\begin{align}
m_s^2 = 12 \frac{m_{3/2}^2}{\Lambda^2}\,,
\end{align}
which, through $m_{3/2}$, depends on $\varphi$ during inflation. The inflaton-dependent minimum of $s$ lies at
\begin{align}\label{eq:smini}
\langle s \rangle =  \frac{2m_{3/2}^2-V_0}{m_{3/2}^2}\frac{\Lambda^2}{4\sqrt{3}}\,.
\end{align}
The scalar potential after integrating out $s$ reads
\begin{align}
V = V_0 \left[ 1 + \left(1-\frac{V_0}{4m_{3/2}^2}\right)\Lambda^2+\mathcal{O}(\Lambda^4)\right]\,.
\end{align}
As mentioned above, the corrections from the sgoldstino sector appear in powers of $m_{3/2}^2/m_s^2$ and $H^2/m_s^2$, where $H\sim\sqrt{V_0}$ again denotes the Hubble parameter. Corrections are under control as long as $m_{3/2} > H \Lambda$ which is the case during inflation in the two examples of Section~\ref{sec:sgoldstinoless}.\footnote{For the sgoldstino to be heavier than $\text{Re}(\Phi)$ one would additionally have to require $|f''(\Phi)|,|f'''(\Phi)|<|f'(\Phi)|$ on the inflationary trajectory. These conditions are, however, already fulfilled by requiring slow-roll inflation.}

Note that even when the corrections are small the sgoldstino can affect post-inflationary cosmology. If the above constraint is violated after inflation, $s$ may no longer trace its minimum. If it gets trapped the associated potential energy can alter late-time cosmology. This is not necessarily problematic and may even induce interesting signatures. We merely point out that decoupling $s$ from all dynamics in the universe requires the bound $m_{3/2} > H \Lambda$ to be satisfied during the entire cosmological evolution. We will show in the following that in a consistent EFT $\Lambda$ cannot be arbitrarily small, making this a very severe constraint. 

%

\section{Corrections from UV completions}

In the previous section we have included corrections to the inflaton potential which arise from the sgoldstino sector. The corrections disappear in the limit $\Lambda\rightarrow 0$ in which the sgoldstino becomes infinitely heavy. But there are more corrections related to the heavy fields living at the scale $\Lambda$. Contrary to the sgoldstino corrections, these scale with $\Lambda^{-1}$ and prevent us from taking the limit $\Lambda\rightarrow 0$. In the following we discuss these UV corrections in two examples. In the first example the sgoldstino mass is generated by Yukawa interactions with heavy fields, in the second example by gauge interactions. Despite their simplicity we expect that our examples are representative of more sophisticated UV embeddings.

%

\subsection{Example 1}

Consider two additional chiral multiplets $X,Y$ with a vector-like mass $M$. We consider $M$ to be large compared to the Hubble scale and the gravitino mass during inflation. We define the model as follows,
\begin{align}
W &= f(\Phi) (1 + \delta  S ) + \lambda S X^2 + M X Y\,,\\
K &=\frac{1}{2} (\Phi+\overline{\Phi})^2 +|S|^2 + |X|^2 + |Y|^2\,.
\end{align}
It bears resemblance to the O'Raifeartaigh model \cite{O'Raifeartaigh:1975pr}. The sgoldstino superfield $S$ obtains a mass term through its coupling to $X$. The parameter $\delta$ is chosen such that the vacuum energy vanishes at the minimum of the potential
\begin{align}
\delta= \sqrt{3} \left( 1 + \frac{2\pi^2 M^2}{\lambda^4}\right) +\mathcal{O}(M^4)\,.
\end{align}
The tree-level scalar potential along the direction ${x=y=0}$ reads
\begin{align}
V = &\ V_0 + \frac{12\pi^2 M^2}{\lambda^4}m_{3/2}^2+ 2\sqrt{3}\left(V_0 - 2 m_{3/2}^2\right)s \nonumber \\ 
 &+ \left(4 V_0^2 - 2m_{3/2}^2\right) s^2 +\mathcal{O}(s^3)\,,
\end{align}
where $V_0 = |f^\prime|^2$ and $m_{3/2}^2\simeq|f|^2$ as before and $\bar{s}=s$ is assumed. The imaginary part of $s$ is stabilized at the origin and does not play a role in our discussion. The tree-level mass of $s$ is negligible compared to the one-loop contribution due to the interaction with $X$. We use the Coleman-Weinberg formula
\begin{align}
V_\text{CW}=\frac{1}{64\pi^2} \text{Str}\,\mathcal M^4 \log\frac{\mathcal M^2}{Q^2}\,,
\end{align}
where $\text{Str}\, \mathcal M^4 = \sum_i(-1)^{2 J_i}(2 J_i+1)m_i^4$ is the trace over the field-dependent mass eigenvalues of states with spin $J_i$. The Coleman-Weinberg potential gives rise to an additional mass term
\begin{align}
V_\text{CW}=\lambda^4\frac{m_{3/2}^2}{\pi^2M^2} s^2+\dots\,.
\end{align}
Note that it is the same as the mass term arising from the equivalent quantum correction to the K\"ahler potential $\Delta K =- |S|^4/\Lambda^2$ with
\begin{align}\label{eq:lambdaM}
\Lambda = \frac{2\sqrt{3}\pi}{\lambda^2}M\,.
\end{align}
We conclude that we obtain the model of Section~\ref{sec:sgold} as a low-energy effective theory and the small shift of $s$ does not affect inflation for sufficiently small $\Lambda$.

However, we have yet to consider the effect of inflation on the sector of heavy fields $X$ and $Y$. Inflation does not induce linear terms for the scalar components $x$ and $y$. However, it generates a bilinear mass term for $x$. The mass of $\text{Im} \,x$ is given by
\begin{align}
m_{\text{Im}\, x}^2 \simeq M^2 - 2 \sqrt{3} \lambda m_{3/2}\,.
\end{align}
Thus, taking the limit $M \to 0$ to make $S$ nilpotent introduces a tachyonic direction in the full theory, which makes inflation impossible.\footnote{We assume $\lambda >0$. In the opposite case $\text{Re}\, x$ is the tachyon.} To obtain a positive squared mass, we obtain the constraint 
\begin{align}
M^2 > 2 \sqrt{3} \lambda m_{3/2}\,.  
\end{align} 
Taking the example of chaotic inflation~\eqref{eq:chaotic} and using $m\simeq 6 \cdot10^{-6}$, $\varphi\sim 15$, this translates into $M > 0.03 \sqrt{\lambda}$. At the same time, to make the sgoldstino sufficiently heavy we have to require that $\Lambda \lesssim 1$ which is equivalent to $M\lesssim 0.09 \lambda^2$, cf.~\eqref{eq:lambdaM}. After combining these two constraints there is a small window at $\lambda \gtrsim 1$ and $M\sim 0.05$, where sgoldstino-less inflation can consistently take place. In this regime the heavy fields remain at their minima and inflation does not receive corrections besides those of Section~\ref{sec:sgold}. Notice, however, that the sgoldstino mass can at most be enhanced by an $\mathcal O(10)$ factor compared to the gravitino mass. This is illustrated in Table~\ref{tab:scales}, where we show a possible choice of scales which leads to successful sgoldstino-less inflation. In the vacuum, the gravitino mass is much smaller than in the inflationary epoch and low-energy supersymmetry can be obtained.

\begin{table}[t]
\centering
\begin{tabular}{ccccc}
\toprule
& $m_{3/2}$ & $m_s$ & $H$ &  $\Lambda$ \\ \noalign{\vskip 1mm}   
\colrule
\noalign{\vskip 1mm}   
inflation & $8\cdot 10^{14}\,\text{GeV}$ & $9\cdot 10^{15}\,\text{GeV}$ & $9\cdot 10^{13}\,\text{GeV}$ &  $7\cdot10^{17}\,\text{GeV}$\\ \noalign{\vskip 1mm}   
vacuum & $10^{5}\,\text{GeV}$ & $10^{6}\,\text{GeV}$ & $\sim 0$ &  $7\cdot10^{17}\,\text{GeV}$ \\
\botrule
\end{tabular}
\caption{Representative example of the different scales appearing in sgoldstino-less inflation. The values during inflation refer to the beginning of observable inflation, 50-60 $e$-folds in the past.}
\label{tab:scales}
\end{table}
%

\subsection{Example 2}

Second, we consider an example where the sgoldstino receives its mass from gauge interactions. We introduce three new chiral multiplets $X$, $Y$, $\Psi$ which carry the charges $q(X)=-1$, $q(Y)=1$ and $q(\Psi)=0$ under a U(1) symmetry.\footnote{In order to avoid anomalies we have to introduce another field $Z$ with charge $q(Z)=-1$. The field $Z$ can be coupled to a new singlet $\Theta$ via a term $Y Z \Theta$ in the superpotential. When $Y$ breaks the U(1) symmetry this becomes a large vector-like mass term for $Z$. In this case $Z$ and $\Theta$ do not affect our analysis, and we neglect them in the following discussion.} We further assume that $q(S)=1$ and define the model by
\begin{align}
W &= f(\Phi) (1 + \delta X S) + \lambda\Psi (X Y -v^2) \,,\\
K &=\frac{1}{2} (\Phi+\overline{\Phi})^2 +|S|^2 + |X|^2 + |Y|^2 + |\Psi|^2\,,
\end{align}
where $\delta$ is again chosen to adjust the cosmological constant. We find
\begin{align}
\delta= \frac{\sqrt{3}}{v}\left(1-\frac{2v^2}{9}+\mathcal O(v^4)\right)\,.
\end{align}
The second term in the superpotential is introduced to break the U(1) symmetry at a high scale. For the same reason as before we set $\bar{x}=x$, $\bar{y}=y$, $\bar{\psi}=\psi$ in the following. The imaginary parts do not play a role in our discussion. Given that ${v\gg \text{Max}(m_{3/2}, H)}$ the U(1) symmetry is broken along the almost $D$- and $F$-flat direction
\begin{align}
x y = v^2\,, \quad s^2-x^2+y^2=0\,, \quad \psi=0\,.
\end{align}
Using these three conditions to eliminate $x$, $y$, and $\psi$ yields the scalar potential
\begin{align}\label{eq:Ex2V}
V&= V_0 + \frac{2}{3}m_{3/2}^2 v^2+ \sqrt{3}\left(2  V_0 - 2 m_{3/2}^2\right)s \nonumber \\ 
 & \;\;\;+ m_s^2 s^2 +\mathcal{O}(s^3)\,,
\end{align}
with 
\begin{align}
m_s^2=\frac{9 m_{3/2}^2}{2 v^2}\,.
\end{align}
This resembles~\eqref{eq:sgoldstinopotential} if we identify $\Lambda=\sqrt{4/3} v$.\footnote{To recover the exact form of~\eqref{eq:sgoldstinopotential} we would have to substitute $m_{3/2}\rightarrow \sqrt{2} m_{3/2}$.} The large mass $m_s$ decouples the sgoldstino and the small shift of $s$ does virtually not affect inflation.

Unfortunately, this is not the end of the story. So far we have worked in the regime $m_{3/2},H \ll v $. We expect additional corrections if either $m_{3/2}$ or $H$ are close to the scale $v$. To find these corrections we must treat $x$, $y$ and $\psi$ as dynamical fields. We perform a Taylor expansion around $s=0$, $\psi=0$, $x=v$, $y=v$ up to second order in the shift of the four fields. Setting the four fields to their new minima, we arrive at the following effective inflaton potential
\begin{align}\label{eq:Ex2V}
V = V_0 -\frac{9}{4}\left(\frac{1}{2g^2}+\frac{1}{\lambda^2}\right) \frac{m_{3/2}^4}{v^4}\,.
\end{align}
Notice the difference to our first example. In this case the shift of the heavy fields during inflation causes a backreaction on the potential. Expression \eqref{eq:Ex2V} only includes the corrections due to the heavy fields. In addition, the sgoldstino-induced corrections of Section~\ref{sec:sgold} arise.

\begin{figure}
\begin{center}
\includegraphics[width=8cm]{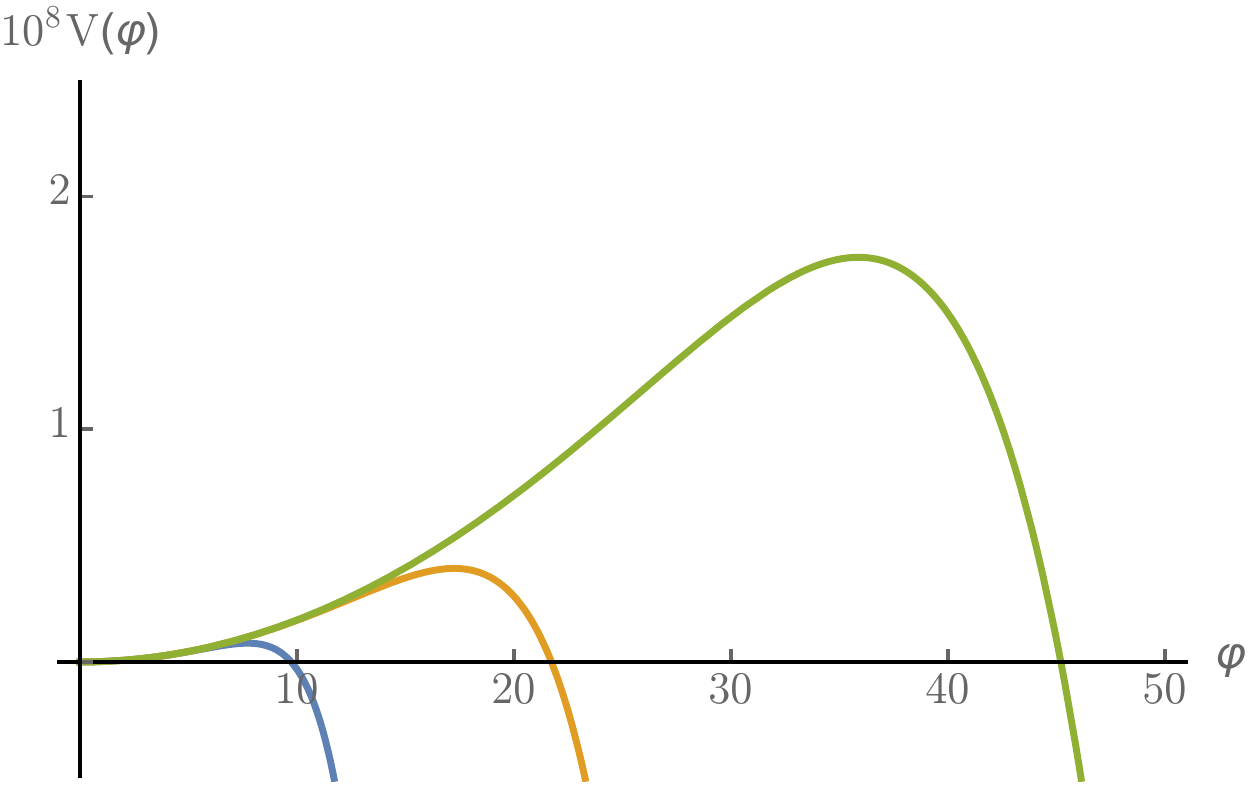}
\caption{Effective inflaton potential for the chaotic inflation model with $v=0.03$ (blue) $v=0.1$ (orange) and $v=0.3$ (green). The backreaction of the heavy fields flattens the inflaton potential. For $v\gtrsim 0.1$ corrections from the heavy fields are under control, while sgoldstino decoupling requires $v\lesssim 1$. This leaves a small window of viable parameter space in which sgoldstino-less inflation consistently proceeds.}
\label{fig:dpotential}
\end{center}
\end{figure}

Requiring the correction to be suppressed compared to the leading-order inflaton potential leads to the constraint
\begin{align}
v\gg \frac{m_{3/2}}{V_0^{1/4}}\,,
\end{align}
for $\lambda, g \sim \mathcal O(1)$. In the model of chaotic inflation defined by~\eqref{eq:chaotic}, with $m\sim 6 \cdot 10^{-6}$ and $\varphi\sim 15$, the constraint translates into
\begin{align}
v \gg 0.03\,.
\end{align}
Even for larger $v$ there are substantial corrections. We depict the effective inflaton potential of the example~\eqref{eq:chaotic} in Figure~\ref{fig:dpotential} for $f_0=10^{-14}$, $m=6 \cdot 10^{-6}$, $\lambda=g=1$, and different values of $v$.
Again there is a small window at $v\gtrsim 0.1$ where the backreaction is under control and sgoldstino-less inflation can take place. As in the previous example, choosing $v$ too large decreases the mass of the sgoldstino scalar beyond the point where it can be consistently decoupled. The corrections from the heavy fields of the UV completion cause a flattening of the inflaton potential. 

%

\section{Discussion}

We have emphasized that sgoldstino decoupling in cosmology is nontrivial. Working in spontaneously broken linear supergravity, instead of imposing a nilpotency constraint by hand we assumed that the mass of the sgoldstino field is produced dynamically. This required the inclusion of heavy degrees of freedom which couple to the sgoldstino. We discussed two possible UV embeddings of nilpotent goldstino multiplets. Both scenarios result in the sgoldstino-less inflation models of \cite{Dall'Agata:2014oka} as a low-energy effective theory. The sgoldstino has a large but finite mass $m_s\sim m_{3/2}/\Lambda$ during inflation, where $\Lambda$ is the mass scale of the heavy fields which couple to the sgoldstino. As $m_{3/2} > H$ the sgoldstino decouples from the inflationary dynamics.

The scale $\Lambda$ sets a new cut-off at which the low-energy effective theory breaks down and the heavy fields become dynamical degrees of freedom. For inflation to take place in a controlled regime, where the heavy fields can be integrated out, one has to require that the Hubble scale does not exceed $\Lambda$. However, an even more severe constraint arises in the class of models~\cite{Dall'Agata:2014oka} which feature $m_{3/2}\gg H$ during inflation. There inflation induces large ``soft terms'' which may destabilize the heavy fields. Depending on the specific UV embedding, we find that tadpole terms $\propto m_{3/2}^2 M_\text{P}^2/\Lambda$ and bilinear terms $\propto m_{3/2}M_\text{P}$ are particularly dangerous. In all examples we find that inflation is generically spoiled if $\Lambda \lesssim 0.1$. This does not leave much room for a complete theory below the Planck scale with a decoupled sgoldstino. Still, a window of viable parameter choices survives in which sgoldstino-less inflation can successfully take place and the backreaction on the heavy fields in under control. Within this window we calculated the corrections to the inflaton potential which typically appear in the form of flattening effects.

The constraints on $\Lambda$ imply that the sgoldstino mass can at most be enhanced by one order of magnitude compared to the gravitino mass. Requiring the sgoldstino to decouple in the post-inflationary cosmology as well puts strong additional constraints on the form of the scalar potential.

Due to the structure of the dangerous terms that arise, we expect these results to be relevant for many other applications of constrained multiplets in cosmology.

%

\section*{Acknowledgements}

The authors thank G.~Dall'Agata and A.~Uranga for discussions. The work of C.W. is supported by the ERC Advanced Grant SPLE under contract ERC- 2012-ADG-20120216-320421, by the grant FPA2012-32828 from the MINECO, and by the grant SEV-2012-0249 of the ``Centro de Excelencia Severo Ochoa'' Programme. The work of L.H. is supported by the IISN and the Belgian Federal Science Policy through the Interuniversity Attraction Pole P7/37  ``Fundamental Interactions". The work of M.W. is supported by the SFB-Transregio TR33 ``The Dark Universe" (Deutsche Forschungsgemeinschaft).

%

\end{document}